\documentclass[11pt]{article}
\usepackage{epsfig}
\usepackage{latexsym,amsmath,amssymb,inputenc,graphicx, float}
\title{Holonomic and non-holonomic geometric models \\ associated to the Gibbs-Helmholtz equation}

\author{Cristina-Liliana Pripoae, Iulia-Elena Hirica, Gabriel-Teodor Pripoae \\ and Vasile Preda}

\date{}
\begin{document}
\maketitle

% Affiliations / Addresses (Add [1] after \address if there is only one affiliation.)
%\address{%
%$^{1}$ \quad Department of Applied Mathematics, The Bucharest University of Economic Studies, Piata Romana 6, \linebreak RO-010374 Bucharest, Romania; cristinapripoae@csie.ase.ro \\
%$^{2}$ \quad Faculty of Mathematics and Computer Science, University of Bucharest, Academiei 14,
%\linebreak RO-010014 Bucharest, Romania; ihirica@fmi.unibuc.ro (I.-E.H.); gpripoae@fmi.unibuc.ro (G.-T.P); preda@fmi.unibuc.ro (V.P.)  \\
%$^{3}$ \quad  ``Gheorghe Mihoc-Caius Iacob'' Institute of Mathematical Statistics
%and Applied Mathematics of Romanian Academy, 2. Calea 13 Septembrie,
%nr.13, sect. 5, RO-050711 Bucharest, Romania \\
%$^{4}$ \quad ``Costin C. Kiritescu'' National Institute of Economic Research of  Romanian Academy, 3. Calea 13 Septembrie, nr.13, sect. 5, RO-050711
%Bucharest, Romania}

% Contact information of the corresponding author
%\corres{Correspondence: gpripoae@yahoo.com}

% Current address and/or shared authorship
%\firstnote{Current address: Affiliation 3.} 
%\secondnote{These authors contributed equally to this work.}
% The commands \thirdnote{} till \eighthnote{} are available for further notes

%\simplesumm{} % Simple summary

%\conference{} % An extended version of a conference paper

% Abstract (Do not insert blank lines, i.e. \\) 
\abstract{ By replacing the internal energy with the free energy, as  coordinates in a "space of observables", we slightly modify (the known three) non-holonomic geometrizations from \cite{Udr2, Udr3, Udr4} and show that the coefficients of the curvature tensor field, of the Ricci tensor field and the scalar curvature function still remain rational functions. \newline
\hskip 10mm In addition, we define and study a new holonomic Riemannian geometric model associated, in a canonical way, to the Gibbs-Helmholtz equation  from Classical Thermodynamics. The main geometric invariants are determined and some of their properties are derived.
Using this geometrization, we characterize the equivalence between the Gibbs-Helmholtz entropy and the  Boltzmann-Gibbs-Shannon, Tsallis and Kaniadakis  entropies, respectively, by means of three stochastic integral equations. We prove that some specific (infinite) families of normal probability distributions are solutions for these equations.}
\medskip

% Keywords

% The fields PACS, MSC, and JEL may be left empty or commented out if not applicable
%\PACS{J0101}
%\MSC{}
%\JEL{}

%%%%%%%%%%%%%%%%%%%%%%%%%%%%%%%%%%%%%%%%%%
%\begin{document}

%%%%%%%%%%%%%%%%%%%%%%%%%%%%%%%%%%%%%%%%%%
%\setcounter{section}{-1} %% Remove this when starting to work on the template.
%\section{How to Use this Template}

%For LaTeX-related questions please contact latex@mdpi.com.%\endnote{This is an endnote.} % To use endnotes, please un-comment \printendnotes below (before References). Only journal Laws uses \footnote.

% The order of the section titles is different for some journals. Please refer to the "Instructions for Authors” on the journal homepage.

\section{Introduction}
\subsection{Motivation}
Classical Thermodynamics is conducted by the Gibbs-Helmholtz (GH) equation, which relates some macroscopic observables (the volume, the free energy (or, alternatively, the internal energy), the pressure, the temperature  and the "entropy") of a closed system. If we interpret it as a Pfaff equation in (an open subset of) ${\mathbb R}^5$, we can consider its kernel, which is a non-integrable (non-holonomic) regular four-dimensional distribution. The non-holonomy forbids the standard (and canonical) application of Riemannian geometric tools on integral (sub)manifolds, so we must appeal for non-holonomic geometrizations. Better than nothing, these non-holonomic tools cannot however catch all the relevant information hidden in the physical model, via the associated distribution.

Our paper has two main goals: firstly, we make a slight variation of three known  Riemanian non-holonomic geometrizations of the GH equation and compare the old and new approaches. Secondly, we avoid the lack of integrability of the previous distribution, by choosing other coordinates. This allows us to consider an holonomic geometrization of the GH equation, which greatly simplifies the framework.

\subsection{History}

At the end of the 19-th Century, the Gibbs-Helmholtz (GH) equation emerged from the papers of J. W. Gibbs and H. Helmholtz and established the rigorous (mathematical) foundation of (Chemical) Thermodynamics. Its interesting story may be read in \cite{Aki, Atk, Ans, Pop, Sag} and in the lively blog of Peter Mander \cite{Man}. The  GH equation is a specific Pfaffian equation, a mathematical notion which was already defined by J.F. Pfaff 100 years before, and involves, among other observables, the so-called "thermodynamic entropy" (a.k.a. "Gibbs-Helmholtz (GH) entropy" or "macroscopic entropy").
\medskip

Approximately at the same time, L. Boltzmann (and soon after M. Planck and J. W. Gibbs) introduced another kind of entropy, suitable for Statistical Mechanics; later, Shannon adapted it for Information Theory. Today, one call it the Boltzmann-Gibbs-Shannon (BGS) entropy (a.k.a. "Gibbs entropy", "Shannon entropy", "information entropy" or "statistical entropy") (\cite{Pop}).
\medskip

Both types of entropy notions have common epistemological roots in Carnot's papers on heat engines at the beginning of 19-th Century and in Clausius work at mid-19-th Century (\cite{Pop}). One century after, their study splitted into two (apparently) divergent theories. Now, \textit{an important open problem is to decide if the two kinds of entropy are equivalent}; in case they are, it would be interesting to establish a "dictionary" between the two theories, and to search for a single Grand Unified Theory of Entropy. This equivalence problem is similar - in some sense- with the equivalence of the inertial and the gravitational mass in the Theory of Relativity (the "Equivalence Principle"). In the (physical, mathematical, epistemological) literature, arguments were brought for both pros and cons variants (equivalence vs. non-equivalence) \cite{Aki, Baw, Cam, Car, Die, Fei, Gau, Guj, Jau, Jay, Kos, Lyn,  Maj, Maro, Mar, Pla, Pru, Pru2, Ser, Swe, Wal, Wer}.

The task to decide where is the truth is all the more difficult as the mathematical methods of approach differ. 
The thermodynamic entropy is a deterministic notion, mainly studied by means of the GH equation, whose modelization is based on contact geometry (\cite{Bad, Bra, Ent, Gei, Gho, Grm, Kho, Kyc, Mru, Sch, Sim} and references therein)  and/or on different non-holonomic associated invariants. The BGS entropy study rests on probability and statistical tools; there exists, however, some geometric objects associated to it, e.g. the Fisher metrics, the statistical manifolds, etc. (see \cite{Hir1, Hir2, Hir3} and references therein),  but  all these notions are of recent birth, when one compares them with the two century old Pfaffian forms; their \textit{long range} relevance and applicability are still to be confirmed.

The Riemannian non-holonomic geometrization  roots in the third decade of the 20-th Century, with the papers of Gh. Vranceanu \cite{Nic1, Vra1, Vra2}  and, independently, of Z. Horak (apud \cite{Kat}). To Pfaffian systems, determining a non-integrable distribution $D$ of interest in Physics (especially in Mechanics), were associated some Riemannian invariants  similar to those from the holonomic known models. Soon after, the theory evolved in many directions, notably in the theory of connections in fiber spaces of E. Cartan and C. Ehresmann.

Through a higher-dimensional analogue of Descartes' trick, 
a complementary orthogonal  distribution $D^{\perp}$ w.r.t. a Riemannian metric $g_D$ establishes an "orthogonal frame" $(D,D^{\perp} )$, which allows a "decomposition" in two parts; the Riemannian machinery 
can be now exploited, producing metric invariants. Given the distribution $D$, there  exist an infinite number of such possible non-holonomic Riemannian models $(D,D^{\perp}, g_D)$ (and many more in the semi-Riemannian setting). The versatility of this approach may be an advantage, but, sometimes, a disadvantage, for both the glory and the limits of non-holonomic geometry. (We avoid entering here in this debate which deserves a better care and a more appropriate framework.)

A highly original geometrization path for dynamical systems, via Pfaffian equations and non-holonomic geometry, is the  Geometric Dynamics of  C. Udriste\cite{Udr1}. In particular, this tool was applied also in the study of the GH equation (\cite{Bad, Udr2, Udr3,Udr4,Udr5,Udr6} to quote but a few).

\subsection{Our contribution}
Our paper deals with three (apparently unrelated) topics: classical thermodynamics and the geometrization of the Gibbs-Helmholtz equation (via holonomic and non-holonomic models); the detalied study of a  hypersurface $\frak S$ in ${\mathbb R}^4$, from both the intrinsic and the extrinsic geometry; the equivalence problem between clasical (thermodynamical) entropy and the statistical entropy. The  unity of the three topics consists in the double role played by the hypersurface $\frak S$: firstly, to prove the advantages of the holonomic approach versus the non-holonomic one; secondly, to be used as a tool for characterizing analytically the (eventual) equivalence between the previous entropy notions.

In Section 2, we recall three (non-holonomic) Riemannian geometrizations of the GH equation, due to Udriste and collaborators. By replacing the internal energy with the free energy, we obtain three new analogous non-holonomic geometrizations, related to the previous ones. The new Riemannian invariants are expressed by rational functions, too.

In Section 3, we make a new {\em holonomic} geometrization of the GH equation, using a special parameterized hypersurface $\frak S$  in ${\mathbb R}^5$. We calculate the matrices of the fundamental forms of this hypersurface, its mean curvatures, its principal curvatures and some of its intrinsic invariants (geodesics, curvature coefficients, Ricci coefficients, scalar curvature). In contrast with the partial/uncomplete tools offered by the non-holonomic models, the geometry of $\frak S$ offers access to the whole Riemannian machinery, which can be used to understand and control the thermodynamic systems.

In Section 4, we use the model from section 3 and we compare the GH entropy with the BGS, the Tsallis and the Kaniadakis entropies, respectively, from Statistical Mechanics. Their equivalence is characterized by  specific  stochastic integral equations. Examples of solutions of these equations are provided. 

We compare our approach with the recent result of Gao et all (\cite{Gao1, Gao2}), which states that (under a set of specific physical assumptions) the BGS (and, eventually, the Tsallis) entropy equals the thermodynamic entropy only for generalized Boltzmann distributions.

In Section 5, we give some thermodynamic interpretation of our results.

\subsection{Conventions}
Some of our definitions and results can be easily extended to deal with  {\em generalized} Gibbs-Helmholtz equations \cite{Ans, Ser, Sag}. We preferred to limit our study  and keep the discourse as elementary as possible, so as not to hide the forest behind the trees.
\medskip

We suppose all the physical quantities  suitably normalized, so that all the equations make sense from the Physics viewpoint.

\section{Avatars of three  non-holonomic Riemannian geometrizations for the GH equation}

Consider a closed thermodynamic system with:   (Gibbs) free energy $G$, pressure $p$,  entropy $S$, temperature $T$, internal energy $U$  and volume $V$. We know that (\cite{Bad, Bla})
%\begin{linenomath}
\begin{equation}
U = G  - pV+TS \quad .
\end{equation}
%\end{linenomath}
The mutual interconnections between these observables are described by the Gibbs-Helmholtz  equation
%\begin{linenomath}
\begin{equation}
dG + S dT - V dp= 0 \quad .
\end{equation}
%\end{linenomath}
Via the relation (1), this equation may be written in the equivalent form

\begin{equation}
dU+ p dV - T dS= 0 \quad .
\end{equation}
 
Define two differential one-forms  $\omega:= dG + S dT - V dp$ and $\eta:= dU+ p dV - T dS $, on two suitable open subsets (as "configurations spaces")  ${\frak D}$ and  ${\frak E}$ in ${\mathbb R}^5$ , respectively, w.r.t. coordinates $(G,p,S,T,V)$ and $(U,p,S,T,V)$.
Then, the equations (2) and (3) can be modeled by the Pfaff equations $\omega=0$ and $\eta=0$, respectively, and by their associated four-dimensional (regular and non-integrable) distributions $ker \omega$ and $ker \eta$. We have

\begin{equation}
ker \omega= span \Big\{ \frac{\partial}{\partial S},  \frac{\partial}{\partial V},  \frac{\partial}{\partial p}+ V \frac{\partial}{\partial G}, \frac{\partial}{\partial T}- S \frac{\partial}{\partial G} \Big\}
\end{equation}
and
\begin{equation}
ker \eta=  span \Big\{  \frac{\partial}{\partial p},  \frac{\partial}{\partial T},  \frac{\partial}{\partial S}+ T \frac{\partial}{\partial U}, \frac{\partial}{\partial V}- p \frac{\partial}{\partial U}\Big\} \quad .
\end{equation}

\noindent{\bf Remark 1.} 
The {\em non-holonomy} of the distribution $ker\omega$ (or, alternatively, $ker\eta$) is the fundamental cause of the difficulty encountered when one tries to integrate the GH equation. For this reason, empirical or more elaborated attempts were invented, and many particular cases were considered, by "slicing" the configuration space or by using idealized models (e.g.  in Carnot-like attempts).
\medskip

From (2)  we obtain

\begin{equation}
S = V \frac{\partial p}{\partial T} - \frac{\partial G}{\partial T} 
\end{equation}
and

\begin{equation}
V = S \frac{\partial T}{\partial p}  + \frac{\partial G}{\partial p} \quad .
\end{equation}

By analogy, from (3) we obtain

\begin{equation}
p = T \frac{\partial S}{\partial V} - \frac{\partial U}{\partial V} 
\end{equation}
and

\begin{equation}
T = p \frac{\partial V}{\partial S}  + \frac{\partial U }{\partial S} \quad .
\end{equation}

\medskip

Udriste and collaborators  used the formalism based on (3) and associated to the  distribution ker $\eta$ three Riemannian metrics  (\cite{Udr2, Udr3, Udr1, Udr4} and references therein), by means of specific techniques  of non-holonomic geometry. They considered global coordinates  $(x^1, x^2, x^3, x^4, x^5):=(U,T,S,p,V) $
and they  determined the respective curvature invariants (Riemann curvature, Ricci curvature and scalar curvature), as rational functions of variables $x^i$, $i=\overline{1,5}$.
\medskip

\noindent{\bf Remark 2.}
An alternative and analogue method is to start from equation (2). W.r.t. the new coordinates $(G,p,S,T,V)$, we can obtain three analogous non-holonomic geometrizations with their corresponding Riemannian invariants. The change of coordinates is non-linear, but involves only rational functions; it follows that the previous curvature invariants are also rational functions, but of variables $G,p,S,T$ and $V$. General covariance laws establish correspondences between the curvature invariants, when calculated in these two systems of coordinates. This simple remark might be important when, in applications, we want to consider the free energy instead of the internal energy of a system. From the theoretical viewpoint, these two formalisms associated to the equivalent forms of the Gibbs-Helmholtz equation (2) and (3) lead to the same geometrizations.
\medskip

\noindent{\bf Remark 3.} Let $\frak V$ be a domain in ${\mathbb R}^3$, of coordinates $(p,G,V)$ (the "configurations space"). Then, the entropy function $S$ in formula (6) can be interpreted as a "Lagrangian" on $\frak V$, i.e. $S : T{\frak V} \to {\mathbb R}$, {\em w.r.t. the temperature} $T$, instead w.r.t. time. We can determine, via formal Euler-Lagrange equations, the "stationary"  curves of the system, of the form
\begin{equation*}
T \rightarrow (p_0,G(T),V_0) \quad ,
\end{equation*}
where both the pressure and the volume are constant. We do not enter this path, because this geometrization is also non-holonomic, even if the non-holonomy is better hidden behind the "velocities space"  $T{\frak V}$. 
\medskip

In the next  section, we leave the realm of non-holonomic geometry and look for geometric properties of thermodynamic systems, with an {\em holonomic} associated model.

\section{A holonomic  geometrization for the GH equation}

With the previous notations, consider $\tilde{p}$ the temperature derivative of the pressure (a.k.a. the thermal pressure coefficient \cite{Abd}) and $\tilde{G}$ the heat (a.k.a. thermal) capacity, i.e. the speed of the free energy w.r.t. $T$. (The notation for the heat capacity is not the usual one !) We can use the formula (6) in order to express the entropy as a function of $\tilde{p}$, $\tilde{G}$ and the volume, i.e. $S=S(\tilde{p},\tilde{G},V)$. Consider coordinates
$(x^1,x^2,x^3):=(\tilde{p},\tilde{G},V)$ on an open subset $\frak{U}$ of ${\mathbb R}^3$.
The entropy function $S(x^1,x^2,x^3)= x^1 x^3-x^2$ on $\frak{U}$ defines a (regular, Monge-type, 3D) hypersurface in ${\mathbb R}^4$. The image of this parameterized hypersurface is a hyperquadric $\frak S$, namely a hyperbolic cylinder in ${\mathbb R}^4$. In Fig.1 one sees how the level sets of $S$ foliate ${\mathbb R}^3$.

\begin{figure}[H]
\includegraphics[width=10.5 cm]{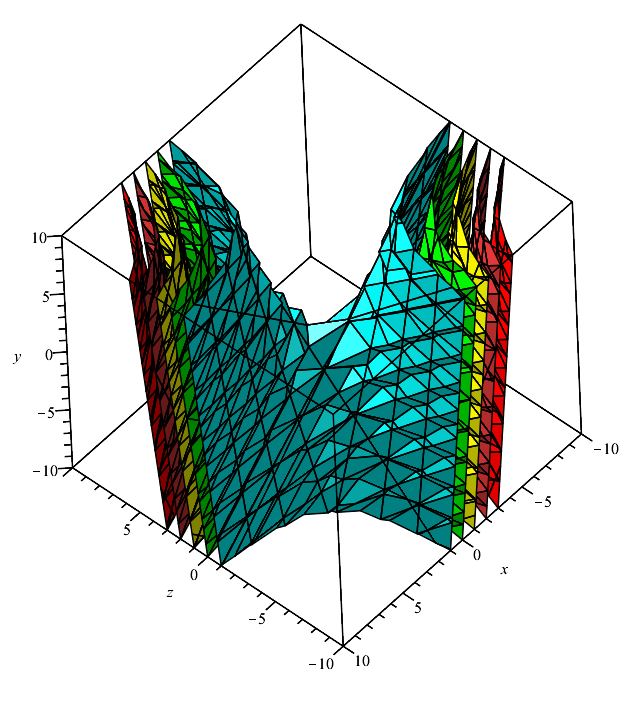}
\caption{The  level sets of $S$.
\label{fig1}}
\end{figure}

The first and the second fundamental forms of  $\frak S$ are, respectively
\begin{equation}
 (g_{ij})_{i,j=1,3}= 
 \begin{pmatrix}
1+(x^3)^2 & -x^3 & x^1 x^3\\
-x^3 & 2 & -x^1\\
x^1 x^3 & -x^1 & 1+(x^1)^2
\end{pmatrix} \quad ,
\end{equation}

\begin{equation}
 (h_{ij})_{i,j=1,3}= a^{-1} \cdot
 \begin{pmatrix}
0 & 0 & -1\\
0 & 0 & 0\\
-1 & 0 & 0
\end{pmatrix} \quad ,
\end{equation}
where $a(x^1,x^3) := \sqrt{2+ (x^1)^2+ (x^3)^2}$. The unit normal vector field is

\begin{equation}
N= a^{-1} \cdot (x^3,-1,x^1,-1) \quad .
\end{equation}

The mean curvature functions of  $\frak S$ are the coefficients of the characteristic polynomial of the second fundamental form w.r.t. the first fundamental form, namely

\begin{equation*}
det(h_{ij}-t g_{ij})= 0  \quad   ,
\end{equation*}
written

\begin{equation*}
t^3 - 3 H_1 t^2+ 3 H_2 t - H_3  =0 \quad   .
\end{equation*}

We calculate:

\begin{equation}
H_1= \frac{6}{a^3} \cdot x^1 x^3 \; , \;  H_2=  -\frac{6}{a^4} \; , \; H_3= 0  \quad   .
\end{equation}

\begin{figure}[H]
\includegraphics[width=9.5 cm]{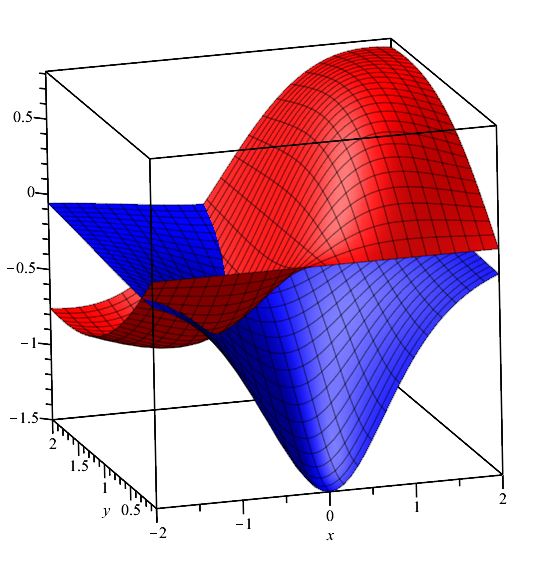}
\caption{The first mean curvature function (red)
and the second mean \\ curvature function (blue). Notation: $x:=x^1$, $y:=x^3$
\label{fig2}}
\end{figure}
We represent graphically, separately, the  first two mean curvature functions, at large scale (only the $x^3 >0$ zone must be retained from the graphics in Fig.3 and Fig.4).

\begin{figure}[H]
\includegraphics[width=10.5 cm]{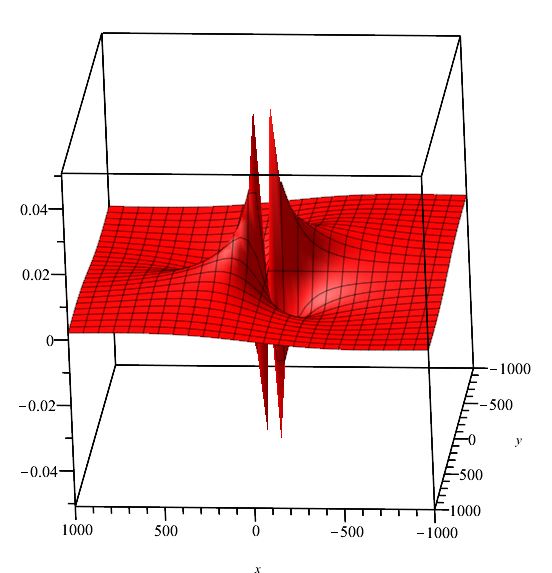}
\caption{The first mean curvature function at large scale. Notation: $x:=x^1$, $y:=x^3$
\label{fig3}}
\end{figure}

\begin{figure}[H]
\includegraphics[width=10.5 cm]{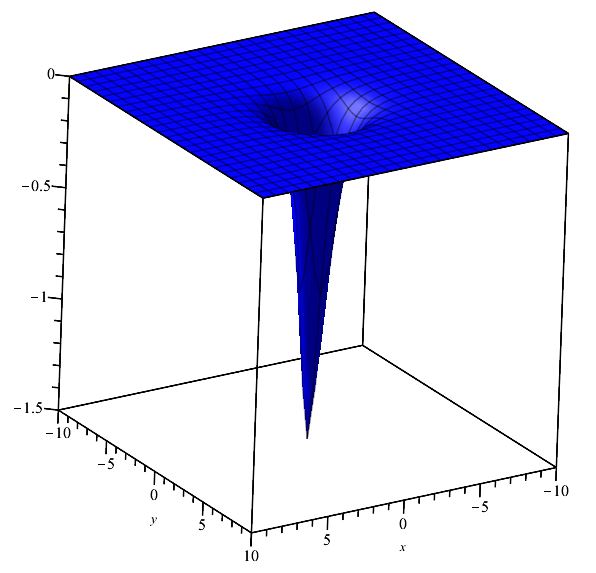}
\caption{The second mean curvature function at large scale. Notation: $x:=x^1$, $y:=x^3$
\label{fig4}}
\end{figure}

The roots of the previous characteristic polynomial are the principal curvature functions of  $\frak S$. We calculate them:

\begin{equation*}
\lambda_1= \frac{1}{a^3} \cdot \Big\{  x^1 x^3 + \sqrt{(2+(x^1)^2)(2+(x^3)^2)}     \Big\} \quad ,
\end{equation*}

\begin{equation*}
 \lambda_2=  \frac{1}{a^3} \cdot \Big\{  x^1 x^3 - \sqrt{(2+(x^1)^2)(2+(x^3)^2)}     \Big\}  \quad ,
\end{equation*}

\begin{equation*}
 \lambda_3=  0  \quad .
\end{equation*}

\begin{figure}[H]
\includegraphics[width=10.5 cm]{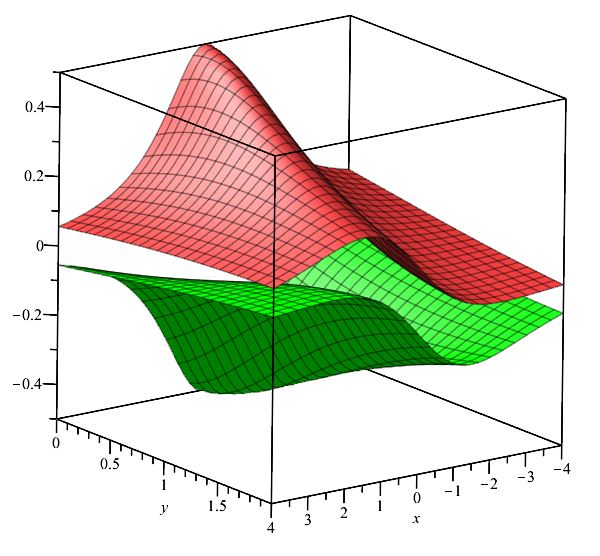}
\caption{The first principal curvature function (orange)
and the \\ second principal curvature function (green). Notation: $x:=x^1$, $y:=x^3$
\label{fig5}}
\end{figure}

\noindent{\bf Proposition 1.} {\it
 The hypersurface $\frak S$ has the following properties: 
\medskip
 
(i) its geometric invariants depend on $x^1$ and $x^3$ only.
\medskip
 
(ii) it is not minimal, totally  geodesic or totally umbilical. Moreover, it has no umbilical points.
\medskip

(iii) it has a null, a positive and a negative smooth principal curvature function. The positive principal curvature function  $\lambda_1 \leq \frac{\sqrt{2}}{2}$, with equality if and only if $x^3=x^1=0$. The negative principal curvature function  $\lambda_2 \geq -\frac{\sqrt{2}}{2}$, with equality if and only if $x^3=x^1=0$. 
\medskip

(iv) it is asymptotically flat.
\medskip

(v) there do not exist extremal values for $H_1$, which is unbounded around $(0,0)$; instead,  $H_2 \leq 0$
and it has a global minimum $-\frac{3}{2}$ at $x^3=x^1=0$.}

\medskip
The intrinsic Riemannian geometry of $\frak S$ can be derived from the first fundamental form only. The (non-null) Christoffel symbols are:

\begin{equation*}
 \Gamma^1_{13}=  \Gamma^1_{31}= \frac{x^3}{a^2} \; , \; 
 \Gamma^2_{13}=  \Gamma^2_{31}= -\frac{1}{a^2} \; , \; 
 \Gamma^3_{13}=  \Gamma^3_{31}= \frac{x^1}{a^2} \quad .
\end{equation*}

\medskip

The  geodesics are solutions of the following ODE system:

\begin{equation*}
\frac{d^2 x^i}{dt^2} + \Gamma^i_{jk} \frac{d x^j}{dt} \frac{d x^k}{dt} =0 \quad , \; i=1,2,3 \quad ,
\end{equation*}
which may be written in detailed form 

\begin{equation}
\frac{d^2 x^1}{dt^2}  +   \frac{2x^3}{a^2} \cdot \frac{d x^1}{dt} \cdot \frac{d x^3}{dt}  = 0 \quad ,
\end{equation}

\begin{equation*}
\frac{d^2 x^2}{dt^2} -   \frac{2}{a^2} \cdot \frac{d x^1}{dt} \cdot \frac{d x^3}{dt}   = 0 \quad ,
\end{equation*}

\begin{equation*}
\frac{d^2 x^3}{dt^2}  +   \frac{2x^1}{a^2} \cdot \frac{d x^1}{dt} \cdot \frac{d x^3}{dt}  = 0 \quad .
\end{equation*}

\medskip

\noindent{\bf Remark 4.}
(i) By contrast with the mean and the principal curvatures formulas, the previous ODEs system depends (formally) on the variable $x^2$.
\medskip

(ii) Numerical solving ODEs system (14), with initial conditions 

\begin{equation*}
x^1(0)= x^2(0)=x^3(0)=1 \; , \; \frac{d x^1}{dt}(0)= \frac{d x^3}{dt}(0)=1 \; , \;  \frac{d x^2}{dt}(0)=10 \quad ,
\end{equation*}
and

\begin{equation*}
x^1(0)= x^2(0)=x^3(0)=1 \; , \; \frac{d x^1}{dt}(0)= \frac{d x^3}{dt}(0)=10 \; , \;  \frac{d x^2}{dt}(0)=1 \quad ,
\end{equation*}
respectively, produces the geodesics in Figure 6 and Figure 7.
\medskip

(iii) As the ODEs system (14) is non-linear, integrating it for exact solutions is a difficult task. We consider only the non-degenerate geodesics.
A general result in global Riemannian geometry assures us that all geodesics are complete. It follows that any two points of $\frak S$ can be joined by a minimizing geodesic.

A first family of geodesics is of the form

\begin{equation*}
x^1(t)= k_1 t + k_2 \; , \; x^2(t)= k_3 t + k_4 \; , \;
x^3(t)=0 \quad ,
\end{equation*}
where $k_1$, $ k_2$, $k_3$ and $ k_4$ are arbitrary constants,
with $ (k_1)^2 + (k_3)^2 \neq 0$.
Another analogous family of geodesics is

\begin{equation*}
x^1(t) =0 \; , \; x^2(t)= k_5 t + k_6 \; , \;
x^3(t)= k_7 t + k_8 \quad ,
\end{equation*}
where $k_5$, $ k_6$, $k_7$ and $ k_8$ are arbitrary constants,
with $ (k_5)^2 + (k_7)^2 \neq 0$.
\medskip

Suppose $x^1=x^1(t)$ and $x^3=x^3(t)$ cannot be null on some open interval of the real line. Then, we have another family of geodesics, with $x^1=x^3$; the function  $x^1$ must satisfy an implicit equation of the form

\begin{equation*}
x^1(t) \sqrt{(x^1(t))^2+1} + ln\Big(x^1(t) + \sqrt{(x^1(t))^2+1}\Big) =k_9 t + k_{10} \quad ,
\end{equation*}
where $k_9$, $ k_{10}$ are arbitrary constants. The second component of the geodesics can be recovered from the second equation in (14), as the anti-derivative

\begin{equation*}
x^2(t) = \int \Big \{ \int \Big[ \frac{1}{(x^1(t))^2+1} \cdot \Big( \frac{d x^1}{dt}(t)\Big)^2 \Big] dt \Big\} dt\quad .
\end{equation*}
The variable $x^2$ will depend on two other arbitrary constants.

The two particular geodesics in Figure 6 and Figure 7 (plotted after numerical integration) belong to this last family.
\medskip

\begin{figure}[H]
\includegraphics[width=10.5 cm]{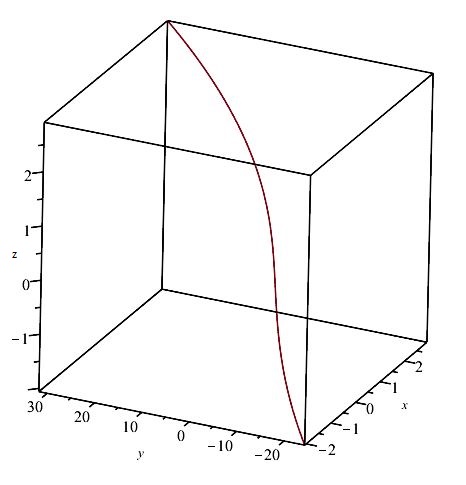}
\caption{The first geodesic. Notation:  $x:=x^1$, $y:=x^2$,  $z:=x^3$
\label{fig6}}
\end{figure}

\begin{figure}[H]
\includegraphics[width=10.5 cm]{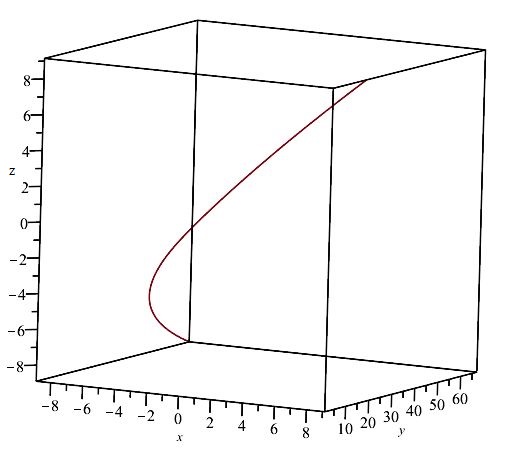}
\caption{The second geodesic. Notation:  $x:=x^1$, $y:=x^2$,  $z:=x^3$
\label{fig7}}
\end{figure}

We  calculate now the (non-null) (0,4)-Riemann curvature coefficients:

\begin{equation*}
  R_{1313}=-R_{1331}= -R_{3113}= R_{3131} = - \frac{1}{a^2} \quad ,
\end{equation*}

%\medskip

\noindent the (non-null) Ricci coefficients: 

\begin{equation*}
Ric_{11} = - \frac{2+(x^3)^2}{a^4} \; , \; Ric_{13}= Ric_{31}= - \frac{x^1x^3}{a^4} \; , \; Ric_{33}= - \frac{2+(x^1)^2}{a^4} \quad ,
\end{equation*}

%\medskip 
 
\noindent and the scalar curvature 

\begin{equation*}
   \rho = - \frac{4}{a^4}  \quad .
\end{equation*}
\medskip

\noindent{\bf Proposition 2.} {\it
The scalar curvature of $\frak S $ is asymptotically flat, and is bounded
$-1 \leq \rho < 0$. Its unique global minimum point is (0,0,0) and
$\rho(0,0,0)= -1$. Moreover, $\rho=  \frac{2}{3} H_2$.}
\medskip

Due to the last property, the graph of the scalar curvature is very similar to the graph of the second mean curvature, and we do not represent it in a separate figure. More interesting seems to be the foliation of $ {\mathbb R}^3$ by its level sets, which are cylinders along the $x^2$ axis.
\medskip

\begin{figure}[H]
\includegraphics[width=10.5 cm]{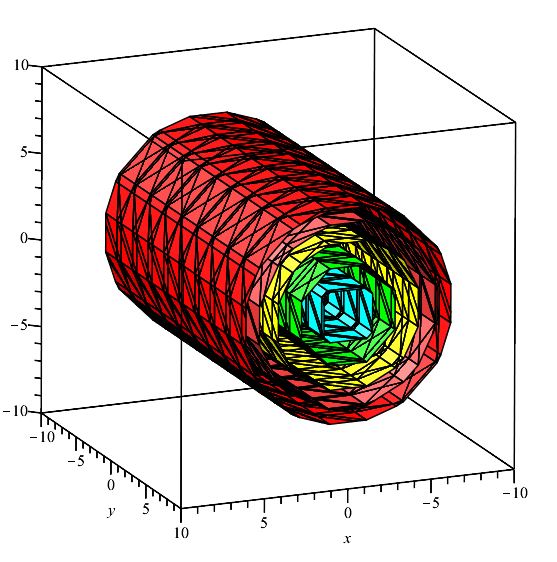}
\caption{The  level sets of $\rho$.
\label{fig8}}
\end{figure}
\medskip

\section{Characterization of the equivalence between the GH entropy and the BGS, the Tsallis and the \\ Kaniadakis  entropy}

Consider a thermodynamical system as in Section 3. Let $M$ be an open set in ${\mathbb R}^n$, $f=f(x,y)$ be a parameterized family of probability distributions (PDFs), $f: {\frak U} \times M \to {\mathbb R}$, with
$\int_{y \in M} f(x,y)dy =1$, $ f \geq 0$.
\medskip

\noindent {\bf Postulate of entropies equivalence.} {\em We suppose that the GH entropy coincides with  the BGS entropy.} (For simpler calculations, the Boltzmann constant is normalized to 1.)
\medskip

This property is characterized by the following {\em equivalence equation}: 

\begin{equation}
 x^1 x^3 - x^2 + \int_{y \in M} f(x,y)\cdot log f(x,y) dy =0 \quad .
\end{equation}
This stochastic  integral equation may be useful when we want to determine an unknown PDF $f$, suitable for a given thermodynamic model. 
\medskip

\noindent{\bf Example 1.} Let 

\begin{equation}
\sigma(x) := \frac{1}{\sqrt{2\pi}} exp \Big\{ x^1 x^3 -x^2 - \frac{1}{2}\Big\}  
\end{equation}
and an arbitrary real valued function $\mu = \mu(x)$, defined on $\frak U$. Consider the family of parametrized normal PDFs on the real line, given by

\begin{equation}
f(x,y):=  \frac{1}{\sqrt{2\pi} \sigma(x)}  exp \Big\{ -\frac{1}{2} \Big( \frac{y-\mu(x)}{\sigma(x)}\Big)^2\Big\}   \quad .
\end{equation}

Then $f$ is a solution of equation (15). We remark that the means may depend arbitrary on the thermodynamic variables. Instead, the dispersion depends inversely proportional on the GH entropy function.
\medskip

Similar solutions of the equation (15) may be looked for w.r.t. other generalized logarithms, instead of the Neperian one. The next two examples use the Tsallis logarithm and the Kaniadakis logarithm, respectively.
\medskip

\noindent{\bf Example 2.} 
We look for solutions for the equivalence equation

\begin{equation}
 x^1 x^3 - x^2 + \int_{y \in M} f(x,y)\cdot log^T_q f(x,y) dy =0 \quad ,
\end{equation}
which is the analogue of equation (15), where the BGS entropy and the Neperian logarithm were replaced by the Tsallis entropy and the Tsallis $q$-logarithm (\cite{Hir3})

\begin{equation*}
log^T_q (z) :=  \frac{z^{1-q}-1}{1-q}  \; , \; q \neq 1 \quad .
\end{equation*}

Suppose $q<2$ and let

\begin{equation}
\sigma(x) := \frac{1}{\sqrt{2\pi}} (2-q)^{\frac{1}{2(q-1)}}
\cdot  \Big[ 1+ (q-1)(x^1 x^3 -x^2) \Big]^{ \frac{1}{q-1}}  \end{equation}
and an arbitrary real valued function $\mu = \mu(x)$, defined on $\frak U$. Consider the family of parametrized normal PDFs on the real line, given by
(17). Then $f$ is a solution of equation (18). We remark that the means may depend arbitrary on the thermodynamic variables. The dispersion depends on the GH entropy function in  more subtle way than in Example 1.

When $q \geq 2$, some integrals become divergent and the previous reasoning does not work anymore. 
\medskip

\noindent{\bf Example 3.} 
We look now for solutions for the equivalence equation

\begin{equation}
 x^1 x^3 - x^2 + \int_{y \in M} f(x,y)\cdot log^K_k f(x,y) dy =0 \quad ,
\end{equation}
which is the analogue of equation (14), where the BGS entropy and the Neperian logarithm were replaced by the Kaniadakis entropy and the Kaniadakis $k$-logarithm (\cite{Hir3})

\begin{equation*}
log^K_k (z) :=  \frac{z^k-z^{-k}}{2k}  \; , \; k \in (-1,1)  \; , \; k \neq 0 \quad .
\end{equation*}
Consider

\begin{equation}
\sigma(x) := \frac{1}{\sqrt{2\pi}} 
\cdot  \Big\{ \frac{k\sqrt{1-k^2}(x^1 x^3 -x^2)+\sqrt{ k^2(1-k^2)(x^1 x^3 -x^2)^2+\sqrt{1-k^2}}}{\sqrt{1+k}} \Big\}^{\frac{1}{k}} 
\end{equation}
and an arbitrary real valued function $\mu = \mu(x)$, defined on $\frak U$. Consider the family of parametrized normal PDFs on the real line, given by
(17). Then $f$ is a solution of equation (20). We remark that the means may depend arbitrary on the thermodynamic variables. The dispersion depends on the GH entropy function, but in a  more complicated way than in Examples 1 and 2.
\medskip

\noindent{\bf Remark 5.}
(i) The previous three examples suggest the following natural question: Which are {\em the familes of} PDFs $F$ (not necessarily normal !) and the generalized "logarithms" $\varphi$ (\cite{Hir3}), such that

\begin{equation}
 x^1 x^3 - x^2 + \int_{y \in M} F(x,y)\cdot \varphi ( F(x,y)) dy =0 \quad ?
\end{equation}

This equation establishes the equivalence of the thermodynamic entropy given by the first two terms and the (statistical) generalized entropy associated to the generalized "logarithm" $\varphi$. Solving it is much more difficult, as the unknowns are  both deterministic ($\varphi$) and stochastic ($F$).

In a previous remark, we explained why we consider only the classical GH equation, and not a generalized one. In the case of {\em generalized} GH equations, the first two terms in (22) are to be replaced by another expression in, eventually, more generalized coordinates (corresponding to more thermodynamic state functions and possibly other statistical quantities). The nature of the problem remains unchanged; all complications arise only as a consequence of the complexity of calculations in a space with more dimensions.
\medskip

(ii) Recently (\cite{Gao1, Gao2}), Gao at all. proved that, under three specific assumptions (of physical inspiration), the only PDF  in which the GBS entropy equals the (classical) thermodynamic entropy is the generalized Boltzmann distribution (i.e. a distribution of  exponential type).  A hint points out that the result may be extended to include the Tsallis entropy as well. This remarkable result gives a partial answer to problem (22).

However, the three assumptions of Gao significantly restrict (from the mathematical perspective) the framework, and weaker hypothesis are desirable. Moreover,  hidden necessary conditions exist behind equation (22), such is the extensivity property; it follows that the  thermodynamic entropy and the statistic entropy (equal to the previous one) must be both extensive or both non-extensive (e.g. for the Tsallis and Kaniadakis entropies \cite{Uma}).
\medskip

(iii) We must make a clarification of terminology. Common language identifies "entropy" as a functional $E=E[f]$ defined of the set of PDFs, with "entropy" as a specific value $E[f_0]$ of this functional. (At a more elementary level, this happens when we speak about "the function $sin t$", instead of "the function $sin$".)

Denote the BGS, the Tsallis and the Kaniadakis entropy functionals with $E^{BGS}$, $E^{T}$, $E^{K}$, respectively. Denote by $f_{BGS}$, $f_T$, $f_K$ the parameterized families of PDFs obtained in the three previous examples. We showed that the thermodynamic entropy $S=S(x)$ coincides (as a function of $x$) with  $E^{BGS}[f_{BGS}(\cdot,y)]$, $E^{T}[f_{T}(\cdot,y)]$ and  $E^{K}[f_{K}(\cdot,y)]$. This does not mean that $S$ (which is a function !) coincides with the functionals (!) $E^{BGS}$, $E^{T}$, $E^{K}$. This is the true meaning of the equivalence  stated in (15) and (22).
\medskip

\noindent{\bf Remark 6.}
Denote $f=f(x,y)$  a family of PDFs on ${\frak U} \times M$, $\varphi$  a generalized logarithm and  

\begin{equation}
 H[f](x) := - \int_{y \in M} F(x,y)\cdot \varphi ( F(x,y)) dy 
\end{equation}
a parameterized family of arbitrary generalized  entropy functionals. Denote $g^{H,f}$ a Riemannian generalized Fisher metric on $\frak U$, canonically associated to $H$ and $f$ \cite{Hir3}.

The thermodynamic entropy $S$ is called {\em metrically equivalent} with the entropy $H[f]$ if the first fundamental form $g$ in (10) coincides with $g^{H,f}$. Variants may include:

\begin{itemize}

\item $g$ and $g^{H,f}$ are homothetic;

\item $g$ and $g^{H,f}$ are conformal;

\item $g$ and $g^{H,f}$ are in geodesic correspondence.

\end{itemize}
The new "equivalence problem"  can now be stated: {\em find $H$ and $f$ such that $S$ be metrically equivalent with $H[f]$}.

This equivalence of entropies in not more general than the previous one in (22), nor an extension or a particularization of it; it is of a different nature, a kind of intermediate equivalence by means of derived objects. The equivalence in (22) and the "metrical equivalence" are logically unrelated.
We do not enter in further details here, as the study requires the whole machinery behind the generalized Fisher metrics \cite{Hir3}.
\medskip

\section{Thermodynamic interpretations and applications}
The previous sections were more mathematical-oriented. Now, we will focus on some physical interpretations of the holonomic model from Section 3 and Section 4.
\medskip

(i) First, we remark  that we use somehow atypical variables, as coordinates for the "space of configurations" $\frak U$ (in addition to the volume $x^3$, which is commonly and frequently used), namely the thermal pressure coefficient $x^1$ and the thermal capacity $x^2$. However, even if these variables/observables are less common in the literature, they are not completely absent (e.g. \cite{Chen, Liu}).
\medskip

(ii) The intrinsic geometry and the extrinsic geometry of the hypersurface $\frak S$ do not depend on the variable $x^2$, so they are independent on the heat capacity $\tilde{G}$. Instead, the set properties of this hypersurface depend on $x^2$.
\medskip

(iii)  Our formalism may be useful when one develops a calculus on the hypersurface  $\frak S$, for example by taking higher order derivatives of
the pressure w.r.t. temperature  (see \cite{Ahl} for second order ones).
\medskip

(iv) Translations can be made between geometric and physical properties: for example, the only points where the first mean curvature function $H_1$ vanishes are the critical points for the pressure function (w.r.t. the temperature).
\medskip

The level sets for the entropy function $S$ (see Fig.1) have deep physical meaning. We look for their intersection with the level sets of the scalar curvature function $\rho$ (see Fig.8), which have interesting mathematical meaning. Namely, let $R \geq \sqrt{2}$ and $a$ a real constant. Consider the 
points   $(x,S(x)) \in {\frak S}$, such that

\begin{equation*}
 S(x) = a \quad , \quad \rho(x) =  - \frac{4}{R^2+2}  \quad .  
\end{equation*}
The intersection curve of the two level sets satisfy the system of two implicit equations and an inequation

\begin{equation*}
 x^1 x^3 - x^2 = a \quad , \quad (x^1)^2 + (x^3)^2 = R^2  \quad , \quad x^3 > 0 \quad . 
\end{equation*}
There exists a unique $\theta \in (0,\pi)$, such that

\begin{equation*}
 x^1 = R cos \theta \quad , \quad x^2 = \frac{1}{2} R^2 sin 2\theta -a \quad , \quad x^3 = R sin \theta \quad . 
\end{equation*}
The parameterized intersection curve $\theta \to (R cos \theta,\frac{1}{2} R^2 sin 2\theta -a, R sin \theta )$
has the  graph in Fig.9.
\begin{figure}[H]
\includegraphics[width=10.5 cm]{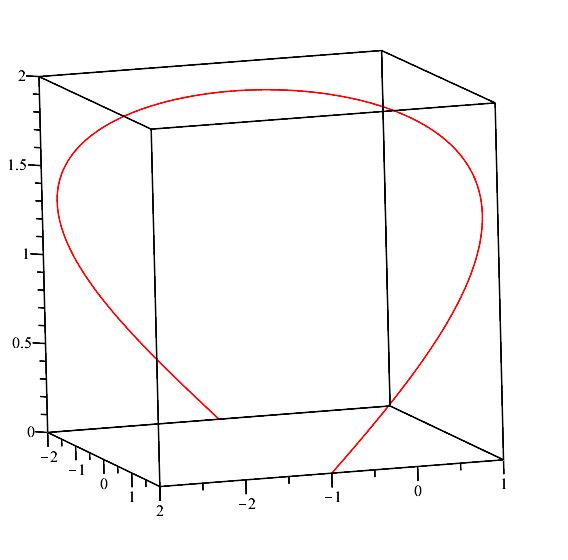}
\caption{The intersection of the two level sets.
\label{fig9}}
\end{figure}
The second coordinate of the intersection curve (which corresponds to the heat capacity $\tilde{G}$ restricted along the intersection curve) suggests a point 

\begin{equation*}
( \frac{1}{2} R^2 cos 2\theta , \frac{1}{2} R^2 sin 2\theta -a) \quad , 
\end{equation*}
situated on a virtual  circle of center $(0,-a)$ and radius $
\frac{R^2}{2} $. Formally, we denote $\tilde{G}^d:= \frac{1}{2} R^2 cos 2\theta$ and call it {\em the mate heat capacity along the intersection curve}. The following formula holds

\begin{equation*}
 (\tilde{G}^d)^2 +  (\tilde{G}^2+a)^2 = \frac{1}{4} R^4  \quad . 
\end{equation*}

We do not know if this quantity can be extended to a (formal, speculative and exotic) new state variable; anyhow, it  has  an interesting intrinsic interpretation.
\medskip

(v) The parameterized PDFs, which arise as solutions of the special stochastic equations in Section 4, are encountered in the literature, in different frameworks (see, for example, \cite{Eln}). Moreover, the geometrization of such parameter spaces lead to the study of statistical manifolds  and of Fisher-like Riemannian metrics in Information geometry (see \cite{Hir1, Hir2, Hir3} and reference therein).
\medskip

(vi) The ODEs system (14) allows the determination of the geodesics lying on the hypersurface $\frak S$. As pointed out in Remark 4 (iii), any geodesic local minimizes the arc length between two points, which can be interpreted as two events in the space of thermodynamic states $x^1$, $x^2$, $x^3$ and $S$. We have here a possible control tool, useful to "drive" a thermodynamic engine from a starting state to a nearby final state. 

More precisely, consider the "state" in $\frak S$ at time $t_0$, characterized by:  $\tilde{p}(t_0)$, $\tilde{G}(t_0)$, $V(t_0)$ and $S(t_0)$. We want to reach the "state" $(A,B,C, S(A,B,C)) \in {\frak S}$, by the "shortest" path. The Remark 4, (iii) ensures us there exists an unique "minimal" geodesic 

\begin{equation}
\gamma= \gamma(t) : [t_0,t_0+b] \to {\frak S} \; , \; \gamma(t_0)= (\tilde{p}(t_0),\tilde{G}(t_0),V(t_0), S(t_0)) \quad ,
\end{equation}
such that $\gamma(t_0+b) = (A,B,C, S(A,B,C))$. Here, "minimal" refers to the Riemannian distance w.r.t. the first fundamental form, not to the Euclidean distance (as the coordinates are not position coordinates). In practice, the geodesic $\gamma$ must be determined numerically, from (14).

Such an approach is, of course, determined/limited by the choice we made, by the particular Riemannian geometry  we found on $\frak S$. There exist other alternative Riemannian metrics with similar claims (\cite{Caf, Sca, Zul}), associated to the GH equation, and a comparison of their practical efficiency and relevance deserves another detailed study.  
\medskip

(vii) The maximum entropy (MaxEnt) problem is a fundamental area of investigation in Statistical Mechanics and Information theory. Its classical thermodynamics counterpart is less studied and, in any case, with totally different tools (\cite{Har}, Ch.5); mathematical optimization with non-holonomic constraints is a difficult theory, which emerged only recently (see \cite{Luc, Udr7, Yos} and references therein).

Our holonomic geometrization allows a direct study, with geometric visualization, of (thermodynamic) entropy fluctuations, including extremum points, on subsets of the hypersurface $\frak S$. 
\medskip

(viii) The geometric model in Section 3 does not take into account the (eventual) positiveness of the entropy. Such an additional condition, if necessary, restricts the framework to  an open set of $\frak U$.
\medskip

(ix) Like other fundamental equations in Physics, the GH equation does not remain valid outside "normal conditions", for example for long range interactions. Our holonomic model in Section 3 can be refined, to cover scale fluctuations. As the coordinates we use are not the "spatial" ones, the Euclidean distance $r$ (as the length of the position vector field in  spherical coordinates) has no longer applicability. We replace the $r$-scale by the $V$-scale, because there is a direct (nonlinear) proportionality between them. 

Let $\nu : (0,\infty) \to (0,\infty)$ be a smooth function, strictly increasing, with the properties:

\begin{equation*}
\lim_{t \to \infty} \nu(t) = \infty \; , \;  \lim_{t \to 0} \nu(t) = 0 \quad .
\end{equation*}

Obviously, there exists a unique $t_0$ such that $\nu(t_0)=1$. Relevant examples are: $\nu(t)= t^{\alpha}$, for a fixed positive $\alpha$;
$\nu(t)= a^{t^b}-1$, for  fixed positive $b$ and $a>1$.

Consider the {\em $\nu$-GH equation}

\begin{equation*}
d G + S^{(\nu)} \cdot dT - \nu(V) \cdot dp = 0 \quad .
\end{equation*}

We derive the formula for the {\em $\nu$-entropy}

\begin{equation*}
 S^{(\nu)}=  \nu(V) \cdot  \tilde{p} - \tilde{G} \quad .
\end{equation*}

In particular, for $\nu = id$, we obtain $S = S^{(id)}$ and we recover formula (6).

By analogy with the computations in Section 3, we obtain a hypersurface ${\frak S}^{(\nu)}$,  we derive a first fundamental form $g^{(\nu)}$, a second fundamental form $h^{(\nu)}$, the mean curvature functions, the principal curvature functions, the scalar curvature function and we can write the equations of the geodesics.

Each member of this infinite family of models "parameterized" by $\nu$ deserves a similar study as those in Section 4 and Section 5. The techniques will be similar but with distinctive outcomes. At "infinity" will dominate the long range interactions with specific (local) entropies; near "zero", for tiny range interactions, we shall obtain different specific entropies.

\section{Discussion}

The first part of the paper contains a short incursion into the realm of non-holonomic geometrizations of GH equations. We did not intend to develop this path, because comparing the possible approaches and further  studies would take too much space. This may be an interesting project for the future. The same remark is valid for an -eventual-  critical study about the pros and the cons of the non-holonomic modelization, when compared to the holonomic one.
\medskip

The results in Section 4 originate in our belief that entropy must be described in a unified way, in Classical Thermodynamics as in Statistical mechanics or Information Theory. We avoided the temptation to postulate it firmly, because we are aware that this hypothesis might look too speculative, from the viewpoint of both theoretical or applied scientists. Our mathematical results are expressed in a neutral approach, leaving open  doors toward unlimited future conclusions.
\medskip

In addition to the content of Sections 4 and 5, more physical interpretations are needed, in order to confirm or to reject our claims. We must investigate if our speculative ideas correspond not only to (possible) "gedanken experiments", but also to  real life thermodynamic systems with significant applications.
\medskip

Developments may include solving the analogue of equations (15), (18) and (20), for other remarkable families of entropies (Renyi, Sharma-Taneja-Mittal, Naudts, etc).
\smallskip

Equations (7), (8) and (9) can be used in order to construct similar holonomic geometrizations of the GH equation. In these cases, one needs a completely different approach to characterize the equivalence of the GH entropy and entropies from Statistical Mechanics (BGS, Tsallis, Kaniadakis, etc). Instead of the "simple" stochastic integral equivalence equations (17), (20, (22), one presumably will obtain more complicated stochastic functional and integral equivalence equations.
\medskip

We restricted our study to the Physics domain, but we must stress there exists another active field of research,  which translates (via a specific dictionary) the thermodynamical notions and results into economic ones \cite{Fer, Geo, Udr4, Udr5, Udr6}. All the content of our paper have a direct correspondence within this economic theory, which remains to be precised and developed in a future paper.
\medskip

In several places in the paper, we emphasized the multitude of Riemannian geometries which can be associated, in various ways, to holonomic or to non-holonomic models for the GH equation. There exist at least two tools to compare any two such geometries. The first one is by means of the deformation algebra associated to the Levi-Civita connections of the respective Riemannian metrics (see \cite{Nic2} and references therein). The second one is  the geodesic correspondence,
which eventually occurs between two Riemannian manifolds and can translate the geodesic dynamics from one space into the other (see, for example, \cite{Nic3}).

%%%%%%%%%%%%%%%%%%%%%%%%%%%%%%%%%%%%%%%%%%
%%%%%%%%%%%%%%%%%%%%%%%%%%%%%%%%%%%%%%%%%%
\vspace{6pt}

%%%%%%%%%%%%%%%%%%%%%%%%%%%%%%%%%%%%%%%%%%

\noindent{\bf Acknowledgments.} The authors dedicate this paper to the memory of their teacher and colleague,  Liviu Constantin Nicolescu (1940-2023), professor emeritus at the University of Bucharest.

%%%%%%%%%%%%%%%%%%%%%%%%%%%%%%%%%%%%%%%%%%
\vskip 1cm

%\begin{adjustwidth}{-\extralength}{0cm}

%\begin{references}


%\bibliography{your_external_BibTeX_file}

%=====================================
% References, variant B: internal bibliography
%=====================================

\bigskip

%%%%%%%%%%%%%%%%%%%%%%%%%%%%%%%%%%%%%%%%%%
Pripoae Cristina-Liliana -  Department of Applied Mathematics, The Bucharest University of Economic Studies, Piata Romana 6,  RO-010374 Bucharest, Romania; cristinapripoae@csie.ase.ro 
\medskip

Hirica Iulia-Elena and Pripoae Gabriel-Teodor - Faculty of Mathematics and Computer Science, University of Bucharest, Academiei 14, RO-010014 Bucharest, Romania; ihirica@fmi.unibuc.ro; gpripoae@fmi.unibuc.ro  
\medskip

Preda Vasile -  "Gheorghe Mihoc-Caius Iacob" Institute of Mathematical Statistics
and Applied Mathematics of Romanian Academy, 2. Calea 13 Septembrie,
nr.13, sect. 5, RO-050711 Bucharest, Romania and
"Costin C. Kiritescu" National Institute of Economic Research of  Romanian Academy, 3. Calea 13 Septembrie, nr.13, sect. 5, RO-050711
Bucharest, Romania, preda@fmi.unibuc.ro 

%%%%%%%%%%%%%%%%%%%%%%%%%%%%%%%%%%%%%%%%%%
\end{document}